\documentclass[10pt]{article}

\usepackage{graphicx}
\usepackage{caption} 
\usepackage{cite}
\usepackage[utf8]{inputenc}
\usepackage[T1]{fontenc}
\usepackage{authblk}
\usepackage{hyperref} 
\usepackage{xcolor} 
\usepackage{amsmath}

\usepackage[left=25mm,top=25mm,bottom=25mm,right=25mm]{geometry}
\title{ {\bf 
Relativistic stars made of anisotropic quark matter in light of current astrophysical constraints: M-R relationship, tidal Love numbers and radial oscillations} 
}

\author{Camila Sepúlveda\thanks{camilaconstanza.sepulveda@ufrontera.cl}}

\author{Grigoris Panotopoulos\thanks{grigorios.panotopoulos@ufrontera.cl}}

\affil{Departamento de Ciencias F{\'i}sicas, Universidad de La Frontera, Casilla 54-D, 4811186 Temuco, Chile.}

\begin{document}

\date{}

\maketitle

\begin{abstract}
We construct interior solutions describing non-rotating, electrically neutral compact stars in Einstein gravity, with the cosmological constant set to zero. The stellar matter is modelled as an anisotropic quark fluid whose equation-of-state is given by the analytic Color-Flavor-Locked (CFL) prescription. We first solve the stellar structure equations and obtain the corresponding mass--radius relation. We then evaluate the gravito-electric tidal Love numbers and the associated dimensionless tidal deformabilities. The resulting configurations are shown to be compatible with current astrophysical bounds arising from massive pulsars, the HESS compact object, NICER observations, and the gravitational-wave events {\bf GW}170817 and {\bf GW}190814. We further examine radial pulsations for three representative stellar masses, calculating the mode frequencies, their eigenfunctions and the large frequency separations. Finally, we compare the three configurations in terms of their radial sound speeds and their fundamental and first-excited oscillation modes.
\end{abstract}

\bigskip


\bigskip

\section{Introduction}

Compact stars offer a natural laboratory for examining the behaviour of matter and gravity in regimes that cannot be reproduced under ordinary terrestrial conditions. White dwarfs, neutron stars and black holes represent possible endpoints of stellar evolution, and the properties of these objects are determined by the extreme densities and gravitational fields involved \cite{Shapiro:1983du}. A reliable account of their structure and dynamics consequently calls for the use of Einstein's General Relativity (GR) \cite{Einstein:1915ca, Psaltis:2008bb}. Relativistic stellar theory can be traced back to Schwarzschild's landmark 1916 solution, which described the exterior space-time associated with a static and spherically symmetric mass distribution \cite{Schwarzschild:1916uq}. Since that time, analytical approaches and numerical methods have been devised to study the equilibrium and observable characteristics of many classes of compact objects. One question of particular relevance is whether the central regions of the most massive neutron stars may contain deconfined quark matter \cite{Alford:2006vz, Weissenborn:2011qu, Baym:2017whm}. Further possibilities arise in models in which compact stars contain additional constituents, including dark matter or dark energy \cite{Grammenos:2016kcq, Sepulveda:2023pvx, Sepulveda:2024eli, Maurya:2024ylr}.

\smallskip

An especially intriguing alternative to an ordinary neutron star is a strange quark star, a hypothetical object consisting largely, or even entirely, of deconfined quark matter \cite{Itoh:1970uw, Collins:1974ky, Haensel:1986qb, Madsen:1998uh, Weber:2004kj}. At sufficiently large densities, the compression of nuclear matter may induce a transition from the hadronic phase to a state in which quarks are no longer confined inside individual nucleons. The strange-matter conjecture makes the stronger assumption that bulk matter containing up, down and strange quarks is absolutely stable, and hence has a lower energy than ordinary nuclear matter \cite{Witten:1984rs, Farhi:1984qu, Dondi:2016yjl, Zhang:2024xod}. Under this hypothesis, strange quark matter would constitute the true ground state of hadronic matter. The possibility that stable bulk matter could instead consist solely of up and down quarks has also been explored under suitable conditions \cite{Benic:2014jia, Holdom:2017gdc}. The interest in such compact configurations is partly motivated by astrophysical objects whose inferred properties are not easily accommodated by standard stellar descriptions \cite{Li:2011zzn, Aziz:2019rgf}; see, in particular, Table 5 of \cite{Weber:2004kj}. In this regard, sources with exceptionally small inferred radii are of special interest \cite{Henderson:2007gu}. A further possible link is with super-luminous supernovae \cite{Ofek:2006vt, Ouyed:2009dr}, whose brightness may be about two orders of magnitude greater than that of conventional supernovae. Despite their rarity, with an estimated rate of approximately one event per thousand supernovae, these explosions provide another motivation for considering alternative channels of compact-object formation.

\smallskip

Any proposed stellar equation-of-state (EoS) should be confronted with a range of independent observational data. The measurements and systems of particular relevance to the present study are outlined below. The gravitational-wave source {\bf GW}190814, observed by LIGO, consisted of a black hole and a companion with an estimated mass in the interval from $2.50 M_{\odot}$ to $2.67 M_{\odot}$ \cite{LIGOScientific:2020zkf}. The classification of this secondary object remains uncertain, since its inferred mass exceeds the range usually attributed to neutron stars. Another noteworthy source is the compact object associated with the supernova remnant HESS J1731-347, for which the reported estimates are a remarkably low mass, $M = (0.77^{+0.20}_{-0.17}) ~ M_\odot$, together with a radius of $R = (10.4^{+0.86}_{-0.78})~km$ \cite{Doroshenko:2022nwp, Rather:2023tly}. These values are of particular interest when compared with the suggested lower limit of $1.17 M_\odot$ for neutron-star masses \cite{Suwa:2018uni}. The binary neutron-star coalescence {\bf GW}170817 constitutes a further source of information, since its gravitational-wave signal yields constraints that include tidal effects \cite{LIGOScientific:2017vwq, LIGOScientific:2018cki}. Independent mass and radius information is supplied by NICER X-ray observations of several pulsars, among them PSR J0030+0451 \cite{Riley:2019yda, Miller:2019cac}, PSR J0437-4715 \cite{Bogdanov:2012md, Choudhury:2024xbk} and PSR J0740+6620 \cite{Riley:2021pdl, Miller:2021qha}. Finally, the measured masses of particularly heavy pulsars, such as PSR J0348+0432 \cite{Antoniadis:2013pzd, Zhao:2015tra} and PSR J1614-2230 \cite{Demorest:2010bx, Lenzi:2012xz}, are close to or exceed twice the solar mass. These observations impose a stringent requirement on the maximum mass attainable by any viable stellar model.

\smallskip

Relativistic stellar models are often developed by treating the material inside the star as a perfect fluid. In this description the pressure is isotropic, meaning that the radial and angular components are identical, $p_r=p_\theta=p_\phi=p$. There is, however, no general requirement that matter under the extreme conditions found in compact stars should obey this restriction. Ruderman discussed the possibility of pressure anisotropy in dense relativistic systems \cite{aniso1}, noting that interactions between relativistic particles may produce distinct radial and tangential stresses. Further theoretical developments of anisotropic stellar matter followed \cite{aniso2}. A number of microscopic or macroscopic processes may give rise to this behaviour, including phase transitions \cite{aniso3}, pion condensation \cite{aniso4} and the presence of type 3A superfluid \cite{aniso5}. Thus, permitting the radial and tangential pressures to differ constitutes a physically motivated generalisation of the conventional perfect-fluid picture.

\smallskip

The inclusion of anisotropic stresses modifies the hydrostatic-equilibrium equations and may lead to appreciable changes in the properties of the resulting stellar configurations. The influence of anisotropy is not confined to the equilibrium structure: it may also affect stellar evolution \cite{Panotopoulos:2025ixx}, mass--radius relations \cite{Wang:1984edr, Benvenuto:1998tx} and tidal responses in binary systems \cite{Madsen:1992sx, Alford:2007pj}. Depending on its sign and magnitude, anisotropy can also make configurations stable that would be unstable in an isotropic treatment \cite{Akmal:1998cf, Andersson:2010sh, Becerra:2024wku}. We define the anisotropic factor through the difference between tangential and radial pressures; with this convention, positive anisotropy corresponds to $p_t>p_r$. Such configurations may support greater maximum masses and larger radii. In appropriate circumstances, the maximum mass can be enhanced by up to $15\%$ compared with the isotropic limit \cite{Becerra:2024wku}. Possible origins of this effect include strong magnetic fields, rotation, and exotic or condensed phases of matter \cite{Becerra:2024wku}. A negative anisotropic factor, on the other hand, corresponds to $p_r>p_t$ and can favour more compact stellar structures. Since these changes may also modify the tidal deformability inferred in binary systems \cite{Panotopoulos:2025ixx, Mohanty:2023hha}, anisotropy should be considered together with both the equilibrium properties and the observational requirements imposed on the model.

\smallskip

For a static, non-rotating relativistic star, hydrostatic equilibrium is described by the Tolman--Oppenheimer--Volkoff (TOV) equations \cite{Oppenheimer:1939ne, Tolman:1939jz}. Given an EoS and appropriate conditions at the centre, numerical integration yields the radial pressure and density distributions, from which the mass, radius and compactness of the star can be determined. The EoS must, of course, be consistent with the assumed composition and microscopic description of the stellar material. The MIT bag model is a commonly employed framework for describing self-bound quark matter \cite{Podder:2023dey}. This approach can be extended to account for colour superconductivity, in particular the Color-Flavor-Locked (CFL) phase, for which an analytic EoS is available \cite{Alford:1998mk, Rajagopal:2000ff, Lugones:2002va, Alford:2007xm}. At sufficiently high density, quarks of different colours and flavours can form Cooper pairs with opposite Fermi momenta \cite{Alford:1998mk, VasquezFlores:2017uor}. In the ideal CFL state, electrical neutrality may be realised without the need for an electron population \cite{Rajagopal:2000ff}. Earlier investigations have shown that colour-superconducting contributions can change the mass--radius relation and permit higher maximum stellar masses \cite{VasquezFlores:2017uor, Lugones:2002zd, Horvath:2004gn}. Moreover, the relativistic framework used to determine stellar structure is not restricted to a single fluid: it can also be applied to systems containing several components. Models combining quark matter with dark matter or dark energy have, for example, been considered in \cite{Sepulveda:2024eli, Grammenos:2016kcq, Panotopoulos:2025ixx, Maurya:2024ylr}.

\smallskip

The inspiral and merger of compact binaries generate gravitational waves that can reveal properties of their constituents which are otherwise challenging to measure. In addition to the component masses, the waveform from a binary containing neutron stars carries information about the response of the stars to the companion's tidal field. The direct observations made by the LIGO \cite{LIGOScientific:2014pky} and Virgo \cite{VIRGO:2014yos} collaborations have consequently established gravitational-wave astronomy as a valuable tool for investigating binaries composed of neutron stars and black holes. The observational prospects are expected to improve substantially with the next generation of detectors. Among the proposed ground-based facilities is the third-generation Einstein Telescope \cite{Punturo:2010zz, ET:2019dnz, Abac:2025saz}; the Laser Interferometer Space Antenna (LISA), in turn, will explore sources in a complementary frequency band \cite{LISA:2017pwj, Barausse:2020rsu}. More accurate observations of compact-binary signals should provide improved opportunities to discriminate between competing descriptions of stellar matter and to constrain the EoS at supranuclear densities.

\smallskip

When two compact objects orbit one another, their mutual gravitational interaction generates tidal fields that deform each star. Since the magnitude of this response is sensitive to the internal density profile, it provides a way of testing the EoS with gravitational-wave data. Within the adiabatic approximation, the response is parametrised by tidal deformabilities and the corresponding Love numbers. Love numbers were originally introduced to describe the deformation of the Earth under an external tidal field, such as that produced by the Moon \cite{Love1, Love2}. The response of an extended body is then encoded in dimensionless coefficients characterising its deformation under an applied gravitational field. For neutron stars and quark stars, the strength of gravity makes a relativistic description necessary; in this setting, the relevant response is determined by perturbations of the space-time metric rather than by the Newtonian gravitational potential alone. Relativistic treatments of tidal effects in spherically symmetric compact objects can be found in \cite{Sepulveda:2023pvx, Flanagan:2007ix, Hinderer:2007mb, Damour:2009vw, Binnington:2009bb, Hinderer:2009ca, Postnikov:2010yn}. The relativistic Love numbers extend the Newtonian coefficient $k$ and relate the solution in the stellar interior to the tidal perturbation in the exterior space-time.

\smallskip

The study of stellar oscillations offers an additional way to probe the properties of stellar interiors. In asteroseismology, the frequencies of pulsation modes are used to infer information about the density distribution, composition and microscopic physics of the stellar material. The method has been applied extensively to the Sun in the form of helioseismology, as well as to other classes of stars, including white dwarfs and red giants \cite{Turck-Chieze:2010rvs, Chaplin:2013hha, Sagun:2020qvc}. Its extension to compact objects has also become an active research topic \cite{Kokkotas:1999bd, Paschalidis:2016vmz, DiClemente:2020szl, Kain:2020zjs}. Studies have addressed neutron stars \cite{Sagun:2020qvc, Rather:2023dom}, strange quark stars \cite{Panotopoulos:2017eig} and dark stars \cite{Sepulveda:2024hdc}. The mode frequencies and their eigenfunctions depend on the equilibrium configuration and on the properties of the matter from which the star is composed. Stellar oscillations may be excited by several astrophysical mechanisms, such as starquakes, tidal interactions during a binary encounter and nearby supernova explosions \cite{Sagun:2020qvc, Franco:1999pma, Hinderer:2016eia, Chirenti:2016xys}. Accordingly, the oscillation spectrum can provide information that complements the constraints derived from mass--radius observations and tidal effects.

\smallskip

Taken together, these theoretical and observational advances make it important to test candidate compact-star models against several classes of data. Compatibility with an individual observation is not by itself sufficient: an acceptable EoS must reproduce the observed range of masses and radii while also satisfying the available limits on tidal deformability. Certain proposed EoSs cannot sustain the largest measured pulsar masses, while others face difficulties in meeting the constraints inferred from gravitational-wave observations. Despite the extensive study of strange quark stars, their viability is determined by both the microscopic EoS and the macroscopic description adopted for the stellar matter. In this work, we analyse a model of anisotropic quark stars, focusing on its consistency with present astrophysical observations and on the oscillation characteristics of configurations that meet the relevant constraints. Our aim is to assess the observational viability of the model and to determine whether its predicted radial-mode spectrum could provide a means of distinguishing it from other viable descriptions.

\smallskip

We focus specifically on strange quark stars whose matter is described by the analytic CFL EoS, while allowing for anisotropic stresses in both the equilibrium and perturbation analyses. As the microscopic quark-matter prescription, we adopt one of the 19 viable models reported in \cite{VasquezFlores:2017uor}, in which colour-superconductivity effects are taken into account. The choice of this model, together with the anisotropy prescription, enables us to confront the resulting stellar configurations with a broad collection of observational constraints. These include NICER measurements, massive pulsars, HESS J1731-347 and the gravitational-wave events {\bf GW}170817 and {\bf GW}190814. We also take into consideration the tidal-deformability bound associated with the improved determination of $\Lambda_{1.4}$ from {\bf GW}170817. Once the compatibility of the selected configurations with these observations has been examined, we turn to their radial oscillations, considering three representative masses, namely $0.7 M_{\odot}$, $2.0 M_{\odot}$ and $2.5 M_{\odot}$. For each stellar model, we calculate the frequencies of the 13 lowest radial modes and study the radial sound speed. We further display the perturbation functions $\eta (x)$ and $\xi (x)$ for the modes $n$ = 0, 1, 5, 6, 11 and 12. This analysis makes it possible to explore how the oscillation properties vary with the stellar configuration and to assess the extent to which radial pulsations can serve as a complementary probe of anisotropic strange quark matter.

\smallskip

The paper is arranged as follows. In Section 2, we present the relativistic equations used to describe hydrostatic equilibrium. Section 3 is devoted to the CFL equation-of-state and to the prescription for the anisotropic factor. The formalism for computing tidal Love numbers, together with the boundary-value problem governing radial stellar oscillations, is discussed in Section 4. Section 5 contains the numerical results and their physical interpretation, whereas the main conclusions are drawn in Section 6. Throughout the paper, we employ the metric signature $(-,+,+,+)$ and work in geometrical units, $G = 1 = c$, where $G$ is Newton's gravitational constant and $c$ is the speed of light in vacuum.

\section{Relativistic fluid spheres within General Relativity}

We now formulate the equations that determine hydrostatic equilibrium for relativistic stars containing anisotropic matter.

\smallskip

We consider non-rotating, four-dimensional fluid spheres with a vanishing cosmological constant. The interior stellar configuration is obtained by solving the equilibrium equations that follow from Einstein's field equations in General Relativity \cite{Einstein:1915ca}:
\begin{equation}
    G_{mn} \equiv R_{mn} - \frac{1}{2} \: R \: g_{mn} = 8 \pi  T_{mn},
\end{equation}

\medskip

\noindent Here, $T_{mn}$ denotes the energy-momentum tensor of the matter source, $g_{mn}$ is the space-time metric, and $R_{mn}$ and $R$ represent the Ricci tensor and Ricci scalar, respectively. The quantity $G_{mn}$ is the Einstein tensor.

\smallskip

For a static and spherically symmetric space-time, the line element written in Schwarzschild-like coordinates $\{ t, r, \theta, \phi \}$ takes the form \cite{Zubairi:2015yda}
\begin{equation}
    d s^2 = - e^{\nu} d t^2 + e^{\lambda} d r^2 + r^2 (d \theta^2 + \sin^2 \theta \: d \phi^2) , 
\end{equation}

\medskip

\noindent We describe the stellar material as an anisotropic fluid, whose energy-momentum tensor is
\begin{equation}
T_a^b = Diag(-\rho, p_r, p_t, p_t),
\end{equation}

\medskip

\noindent In this expression, $\rho$ denotes the energy density, whereas $p_r$ and $p_t$ are, respectively, the radial and tangential pressures.

\smallskip

The stellar mass $M$ and radius $R$ are determined by integrating the Tolman-Oppenheimer-Volkoff (TOV) equations \cite{Oppenheimer:1939ne, Tolman:1939jz}, which describe the interior equilibrium configuration. For a single fluid component with a non-zero anisotropic factor, the corresponding structure equations read \cite{Bowers:1974tgi}
\begin{align}
    \displaystyle m'(r) &= 4 \pi r^2 \rho (r) , \\ 
    \displaystyle \nu' (r) &= 2 \: \frac{m(r) + 4 \pi r^3 p_r(r)}{ r^2 \left( 1 - 2 m(r) / r \right) } , \\
    \displaystyle p_r'(r) &= - [ \rho(r) + p_r(r) ] \; \frac{\nu' (r)}{2} + \frac{2 \: \Delta (r)}{r} ,
\end{align}

\smallskip

\noindent A prime indicates differentiation with respect to the radial coordinate $r$. The mass function $m(r)$ is introduced through the metric component
\begin{equation}
    \displaystyle e^{\lambda} = \frac{1}{1 - \frac{2 \: m(r)}{r}},
    \label{eq:2}
\end{equation}
while the anisotropic factor is defined as $\Delta (r)=p_t-p_r$.

\smallskip

The coupled differential equations are integrated from the centre to the stellar surface. This requires regularity conditions at the origin ($r = 0$), together with appropriate matching conditions at the boundary ($r = R$). The central conditions are
\begin{equation}
    m(0) = 0 ,
\end{equation}
\begin{equation}
p_r(0) = p_t(0) = p_c ,
\end{equation}

\medskip

\noindent where $p_{c}$ denotes the pressure at the centre of the star. At the outer boundary, the interior geometry is matched to the vacuum Schwarzschild solution \cite{Schwarzschild:1916uq},
\begin{equation}
d s^2 = -f(r) d t^2 + f(r)^{-1} dr^2 + r^2 (d \theta^2 + \sin^2 \theta \: d \phi^2), \; \; \; \; \; \; \; f(r) = 1-2 \frac{M}{r} .
\end{equation}

The resulting surface matching conditions are
\begin{equation}
p_r(R) = 0,
\end{equation}
\begin{equation}
m(R) = M, 
\end{equation}
\begin{equation}
e^{\nu(R)} = 1-2 \frac{M}{R} .
\end{equation}

\medskip

The first two relations determine the stellar radius and total mass. The remaining metric potential, $\nu(r)$, can then be obtained from
\begin{equation}
    \displaystyle \nu (r) = \ln \left( 1 - \frac{2 M}{R} \right) + 2 \int_R^r \frac{m(x) + 4 \pi x^3 p(x)}{ x^2 \left( 1 - 2 m(x) / x \right) } \: dx .
    \label{eq:3}
\end{equation}

\section{Matter content}

This section specifies the material description adopted in our stellar models. We first introduce the phenomenological prescription used for the anisotropic factor and then present the relation connecting the energy density with the radial pressure.

\subsection{Anisotropic factor}

For a static, spherically symmetric relativistic star composed of anisotropic matter, the system contains five unknown functions: two associated with the gravitational field, $m(r)$ and $\nu(r)$, and three describing the matter distribution, namely $\rho(r)$, $p_r(r)$ and $\Delta(r)$. Einstein's equations, however, provide only three independent relations. Consequently, two further conditions are required to close the system. A commonly adopted and straightforward phenomenological choice is to specify an equation-of-state (EoS), $\rho(p_r)$, together with an ansatz for $\Delta$. This separation is natural because the anisotropy $\Delta$ is not a thermodynamic variable, and hence the EoS does not, by itself, determine the tangential pressure $p_t$. Furthermore, the third TOV equation governs the radial variation of $p_r$, rather than that of $p_t$. It is therefore appropriate to prescribe the functions $\rho(p_r)$ and $\Delta(p_r)$ independently. In what follows, we adopt precisely this standard procedure: an EoS is specified and a separate functional form is assigned to the anisotropic factor.

\smallskip

At present, there is no unique microscopic theory that provides a model-independent prediction for the anisotropic stress tensor of dense quark matter. Nevertheless, several prescriptions have been developed and employed in the study of compact stars. Examples include the Karmarkar condition on the gravitational sector \cite{Karmarkar}, Ovalle's gravitational decoupling approach \cite{Ovalle1,Ovalle2}, and Herrera's vanishing-complexity-factor condition \cite{Herrera}. Here we instead adopt a phenomenological description. Following \cite{Horvat, Folomeev, Arbanil, Rincon}, we take the anisotropic factor to be given by the Horvat ansatz
\begin{equation}
    \displaystyle \Delta (r) = g \: p_r(r) \: [1-e^{-\lambda(r)}] = \frac{2 \: g \: m(r) \: p_r(r)}{r} ,
\end{equation}

\medskip

\begin{figure}[t]
    \centering 
    \includegraphics[scale=1]{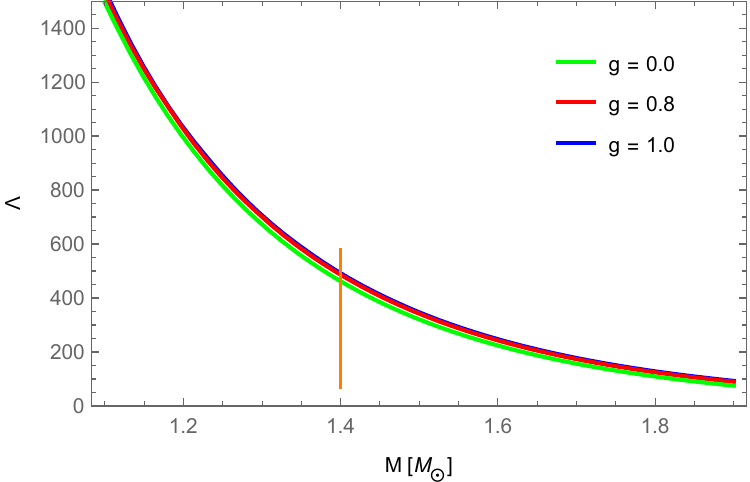}
    \caption{
    Dimensionless deformability $\Lambda$ of our models ({\bf in green for $g = 0.0$}, in red for $g = 0.8$ and blue for $g = 1.0$). The range of values of $\Lambda$ for the gravitational wave signal {\bf GW}170817 is indicated in orange. This means that for a mass of 1.4 $M_{\odot}$, the deformability will have a range of values of $\Lambda_{1.4} \: = \: 190^{+390}_{-120}$ \cite{LIGOScientific:2018cki}.}
    \label{fig:lambda}
\end{figure}

\noindent In this expression, $g$ is a dimensionless parameter governing the magnitude of the anisotropy; the isotropic limit is recovered by setting $g=0$. The Horvat prescription is a convenient closure relation for the otherwise underdetermined system, as it introduces just one additional free parameter and permits a direct comparison with the isotropic case. In addition, the chosen form of $\Delta$ is regular at the stellar centre and vanishes at the surface. It also becomes negligible in the Newtonian regime, where the local compactness, $2 m(r)/r$, is very small. In the analysis below, we consider the two values $g = 0.8, 1.0$ for the dimensionless anisotropy parameter, both of which correspond to a positive anisotropic factor. These values are selected for two reasons. On the one hand, being of order unity, they do not represent either an exceptionally weak or an excessively strong anisotropy. On the other hand, they allow the model under consideration to yield, simultaneously, a mass--radius relation and tidal deformabilities compatible with a broad range of current astronomical observations. The corresponding results are presented in Figure \ref{fig:profile} and Figure \ref{fig:lambda}, and are discussed below.

\subsection{Equation-of-state}

The EoS must be specified before the TOV system can be integrated. In the case of strange quark stars, we employ the MIT bag model and adopt the Color-Flavor-Locked (CFL) equation-of-state.

\smallskip

In terms of the quark chemical potential, the radial pressure $p_r(\mu)$ and energy density $\rho(\mu)$ are given, respectively, by \cite{Lugones:2002va}
\begin{equation}
    \displaystyle p_r (\mu) = \frac{3 \mu^4}{4 \pi^2} + \frac{9 \alpha \mu^2}{2 \pi^2} - B ,
    \label{eq:1}
\end{equation}
\begin{equation}
    \displaystyle \rho (\mu) = \frac{9 \mu^4}{4 \pi^2} + \frac{9 \alpha \mu^2}{2 \pi^2} + B ,
\end{equation}

\medskip

\noindent Here, $\mu$ denotes the quark chemical potential, $B$ is the bag constant, and the parameter $\alpha$ is defined as
\begin{equation}
    \displaystyle \alpha = \frac{2 \Delta_0^2}{3} - \frac{m_s^2}{6} ,
\end{equation}

\medskip

\noindent where $\Delta_0$ denotes the superconducting gap and $m_s$ is the mass of the strange quark.

\medskip

The energy density can be recast as a function of the radial pressure in the form
\begin{equation}
    \displaystyle \rho (p_r) = 3 p_r (r) + 4 B - \frac{9 \alpha \mu^2}{\pi^2} ,
\end{equation}
\noindent where
\begin{equation}
    \displaystyle \mu^2 = \left[ \frac{4}{3} \pi^2 (p_r (r) + B) + 9 \alpha^2 \right]^{1/2} - 3 \alpha .
\end{equation}

\medskip

One may equivalently express the radial pressure in terms of the {\bf energy density}, rather than writing the EoS as $\rho(p_r)$. The resulting expression for $p_r(\rho)$ is
\begin{equation}
    \displaystyle p_r(\rho) = \frac{\rho (r)}{3} - \frac{4 B}{3} + \frac{3 \alpha \mu^2}{\pi^2} ,
\end{equation}
\noindent with
\begin{equation}
    \displaystyle \mu^2 = \left[ \frac{4}{9} \pi^2 (\rho (r) - B) + \alpha^2 \right]^{1/2} - \alpha.
\end{equation}

\medskip

The parameters $\Delta_0$ and $B$ are subject to physical restrictions and cannot be assigned arbitrary values. Within the MIT bag model, these restrictions take the form of two conditions that define the so-called stability window \cite{VasquezFlores:2017uor}. The first condition requires the energy per baryon of CFL quark matter at zero pressure, $p = 0$, and zero temperature, $T = 0$, to be smaller than the neutron mass, $m_n = 939 \: \mathrm{MeV}$. Since this energy per baryon is equal to $3 \mu$, the condition reads \cite{VasquezFlores:2017uor, Farhi:1984qu}
\begin{equation}
    3 \: \mu \leq m_n.
\end{equation}

\medskip

This inequality follows from the fact that all three flavours in CFL quark matter share the same Fermi momenta.

\smallskip

The second requirement states that, at zero pressure and zero temperature ($p=0, T = 0$), the energy per baryon of a pure gas of up and down quarks must exceed the neutron mass, $m_n = 939 \: \mathrm{MeV}$. This requirement places a lower bound on the MIT bag constant, namely $B > 57 \: \mathrm{MeV \: fm}^{-3}$ \cite{VasquezFlores:2017uor, Farhi:1984qu}. Furthermore, for a specified value of $\Delta_0$, setting the pressure in Equation \eqref{eq:1} to zero, $p_r = 0$, determines the corresponding upper bound on $B$. Thus, the allowed interval for the bag constant is
\begin{equation}
    \displaystyle 57 \: \mathrm{MeV \: fm}^{-3} < B < \frac{\Delta_0^2 m_n^2}{3 \pi^2} + \frac{m_n^4}{108 \pi^2} - \frac{m_s^2 m_n^2}{12 \pi^2}.
\end{equation}

For the particular model examined here, we set $B = 80 \: \mathrm{MeV \: fm}^{-3}$, $\Delta_0 = 150 \: \mathrm{MeV}$ and $m_s = 150 \: \mathrm{MeV}$. This parameter choice corresponds to model CFL10 in \cite{VasquezFlores:2017uor}. We note that the 2026 Particle Data Group Review \cite{PDG2026} quotes a strange-quark mass of $m_s(Q=2~GeV) \approx 93~MeV$. This value is a running mass and therefore depends on the renormalization scale $Q$. It should not be confused with the constituent strange-quark mass used as an input parameter in phenomenological models of hadrons. Moreover, since our purpose is to investigate how anisotropy affects quark-star modelling, we keep the parameters of the CFL EoS fixed throughout the analysis. Their variation has already been explored in earlier studies; see, for instance, \cite{VasquezFlores:2017uor, Roupas:2020nua}.

\bigskip

\section{Equations for perturbations}

This section presents the formalism used to describe tidal Love numbers in relativistic stars, together with the equations governing the perturbations of pulsating stellar configurations.

\subsection{Gravito-electric tidal Love numbers}

\hspace{0.5cm}

A detailed account of the relativistic formulation of tidal Love numbers for compact stars is provided, for example, in \cite{Flanagan:2007ix, Hinderer:2007mb, Damour:2009vw, Binnington:2009bb, Hinderer:2009ca, Postnikov:2010yn}. Consider a star immersed in an external gravitational field, $\Phi_{ext}$, which may, for instance, be generated by a companion object in a binary system. In response to this external field, the star undergoes a deformation. At leading order, this deformation gives rise to a quadrupole moment $Q_{ij}$
\begin{equation}
Q_{ij} = \int d^3x \delta \rho(\vec{x}) \: (x_i x_j - \frac{1}{3} r^2 \delta_{ij}),
\end{equation}

\medskip

\noindent which is proportional to the static external quadrupolar tidal field $E_{ij}$
\begin{equation}
Q_{ij} = - \lambda \: E_{ij},
\end{equation}
\begin{equation}
E_{ij} = \frac{\partial^2 \Phi_{ext}}{\partial x^i \partial x^j},
\end{equation}

\medskip

\noindent with the spatial indices taking the values $i,j=1,2,3$.

\smallskip

The dimensionless quadrupolar tidal Love number $k_2$ encodes information about the stellar structure and consequently depends on the mass and equation of state of the star. It is directly connected to two related quantities, usually termed deformabilities: the dimensionful parameter $\lambda_2$ and its dimensionless counterpart $\Lambda_2$, defined by
\begin{align}
    \lambda_2 &\equiv \frac{2}{3} k_2 R^5,
    \\
    \Lambda_2 &\equiv \frac{2 k_2}{3 C^5},
\end{align}

\medskip

\noindent where $C=M/R$ denotes the stellar compactness. The Love number may then be expressed as a function of $C$ in the form \cite{Flanagan:2007ix, Hinderer:2007mb, Damour:2009vw, Binnington:2009bb, Hinderer:2009ca, Postnikov:2010yn}
\begin{align}
    k_2 &= \frac{8C^5}{5} \: \frac{K_{o}}{3  \:K_{o} \: \ln(1-2C) + P_5(C)} ,
    \label{elove}
    \\
    K_{o} &= (1-2C)^2 \: [2 C (y_R-1)-y_R+2] ,
    \\
    y_R &\equiv y(r=R) ,
\end{align}
where $P_5(C)$ is a polynomial of fifth order, explicitly given by
\begin{equation}
\begin{split}
P_5(C) = \: & 2 C \: \Bigl( 4 C^4 (y_R+1) + 2 C^3 (3 y_R-2) \ + 
\\
&
2 C^2 (13-11 y_R) + 3 C (5 y_R-8) -
\\
&
3 y_R + 6 \Bigl)  ,
\end{split}
\end{equation}
%
%
The function $y(r)$ is obtained by solving the Riccati differential equation \cite{Postnikov:2010yn}:
%
\\
\begin{equation}
    r y'(r) + y(r)^2 + y(r) e^{\lambda (r)} \Bigl[1 + 4 \pi r^2 ( p_r(r) - \rho (r) ) \Bigl] + r^2 Q(r) = 0 ,
\end{equation}

\medskip

\noindent with the central initial condition $y(0)=2$ as $r \rightarrow 0$. The function $Q(r)$, which is distinct from the tensor $Q_{ij}$ introduced above, is defined as \cite{Biswas:2019gkw}

\begin{equation}
    \displaystyle Q(r) = 4 \pi e^{\lambda (r)} \Bigg[ 4 \rho (r) + 4 p_r(r) + 4 p_t(r) + \frac{\rho (r) + p_r(r)}{A(r) \: c_r(r)^2} [1+c_r(r)^2] \Bigg] - 6 \frac{e^{\lambda (r)}}{r^2} - \Bigl[\nu' (r)\Bigl]^2,
\end{equation}

\medskip

\noindent where $A=c_t^2/c_r^2$, and $c_r$ and $c_t$ denote, respectively, the radial and tangential sound speeds. These are calculated according to
\begin{equation}
c_r^2(r) = \frac{p_r'(r)}{\rho'(r)}, \qquad c_t^2(r) = \frac{p_t'(r)}{\rho'(r)} .
\end{equation}
where the prime denotes differentiation with respect to $r$. In the isotropic limit, the radial and tangential pressures coincide, as do the corresponding sound speeds: $p_r=p_t=p$, $c_t=c_r=c_s$, and $A=1$. Under these conditions, $Q(r)$ reduces to its standard expression
\begin{equation}
    \displaystyle Q(r) = 4 \pi e^{\lambda (r)} \Bigg[ 5 \rho (r) + 9 p(r) + \frac{\rho (r) + p(r)}{c^2_s(r)} \Bigg] - 6 \frac{e^{\lambda (r)}}{r^2} - \Bigl[\nu' (r)\Bigl]^2 ,
\end{equation}

\medskip

\noindent where $c_s^2 \equiv dp/d\rho$ is the common squared sound speed in the radial and tangential directions.

\smallskip

As discussed in \cite{Damour:2009vw, Hinderer:2009ca, Postnikov:2010yn}, the above expressions require a correction in the presence of phase transitions or discontinuities in the energy density. In the present analysis, the energy density remains finite and non-zero at the stellar surface; we therefore include the following correction:
\begin{equation}
y_R \rightarrow y_R - 3 \frac{\Delta \rho}{\langle \rho \rangle},
\end{equation}
where $\Delta \rho$ denotes the density discontinuity, while
\begin{equation}
\langle \rho \rangle = \frac{3M}{4 \pi R^3},
\end{equation}

\medskip

\noindent represents the mean energy density of the star.

\smallskip

Since $k_2 \propto (1-2C)^2$, it follows that the tidal Love numbers of black holes vanish: their compactness is $C=1/2$. This is a noteworthy prediction of classical GR, according to which black holes, unlike other classes of compact objects, possess exactly zero tidal Love numbers. Consequently, the detection of a non-zero value of $k_2$ would provide a distinctive signature of a departure from the standard black-hole solution of GR. In the context of gravitational-wave astronomy, the Einstein Telescope is expected to determine the neutron-star EoS with high precision \cite{Iacovelli:2023nbv}. The Laser Interferometer Space Antenna, in turn, can investigate objects with very large compactness ($C > 1/3$) and is expected to constrain the Love numbers of highly spinning central objects to a level of approximately $\sim 0.001-0.01$ \cite{Piovano:2022ojl}.

\bigskip

\subsection{Radial oscillations of pulsating stars}

\hspace{0.5cm}

For a spherically symmetric configuration in which the motion is purely radial, Einstein's field equations provide a framework for determining the radial oscillation properties of a static equilibrium star \cite{Chandrasekhar:1964zz, Chandrasekhar:1964zza, Kokkotas:2000up}. The stellar radial perturbations are described by the variables \cite{chanmugam1, chanmugam2}
\begin{equation}
    \displaystyle \xi \equiv \frac{\Delta r}{r},
\end{equation}
\begin{equation}
    \displaystyle \eta \equiv \frac{\Delta p}{p}, 
\end{equation}

\medskip

\noindent where $\Delta r$ denotes the radial displacement and $\Delta p$ the perturbation in pressure. These variables obey the following coupled first-order differential equations \cite{chanmugam1, chanmugam2}
\begin{align}
    \displaystyle \xi'(r) &= - \left( \frac{3}{r} + \frac{p'}{\zeta} \right) \xi - \frac{1}{r \Gamma} \eta , \\
    \displaystyle \eta'(r) &= \omega^2 \left[ r \left( 1 + \frac{\rho}{p} \right) e^{\lambda - \nu} \right] \xi - \left[ \frac{4 p'}{p} + 8 \pi \zeta r e^{\lambda} - \frac{r (p')^2}{p \zeta} \right] \xi - \left[ \frac{\rho p'}{p \zeta} + 4 \pi \zeta r e^{\lambda} \right] \eta , 
\end{align}

\medskip

\noindent Here, $e^{\lambda}$ and $e^{\nu}$ are the metric potentials appearing in Eqs. \eqref{eq:2} and \eqref{eq:3}, respectively, and $\zeta$ is defined as
\begin{equation}
    \displaystyle \zeta \equiv p + \rho . 
\end{equation}

\medskip

The quantity $\Gamma$ denotes the relativistic adiabatic index and is given by
\begin{equation}
    \Gamma = c_s^2 \left( 1 + \frac{\rho}{p} \right) ,
\end{equation}

\medskip

\noindent where $c_s^{2}$ is the squared sound speed, defined through
\begin{equation}
    c_s^2 = \frac{d p}{d \rho} = \frac{p'(r)}{\rho'(r)}. 
\end{equation}

\medskip

The perturbation equations written above describe the radial oscillation modes of stars composed of isotropic matter. For anisotropic configurations, additional contributions proportional to the anisotropic factor arise. The resulting linear system of coupled perturbation equations is \cite{Arbanil}
\begin{equation}\label{ksi}
     \xi'(r) = -\frac{1}{r} \Biggl( 3\xi + \frac{\eta}{\Gamma} \Biggr) - \left( \frac{P'(r)}{P+\rho} + \frac{2 \Delta}{r P \Gamma} \right) \xi(r),
\end{equation}
\begin{equation}\label{eta}
    \begin{split}
          \eta'(r) = \xi \Biggl[ \omega^{2} r (1+\rho/P) e^{\lambda - \nu } - \frac{4P'(r)}{P} -8\pi (P+\rho) re^{\lambda} \frac{P+\Delta}{P} \\
     +  \frac{8 \Delta}{r P} + \frac{r(P'(r))^{2}}{P(P+\rho)}\Biggr] + \eta \Biggl[ -\frac{\rho P'(r)}{P(P+\rho)} -4\pi (P+\rho) re^{\lambda}\Biggr] ,
    \end{split}
\end{equation}

\medskip

\noindent In these equations, $P$ and $\Gamma$ refer to the radial quantities, namely $P(r)=p_r(r)$ and $\Gamma(r)=\Gamma_r(r)$.

\smallskip

The derivation of the radial-stability equations can be outlined as follows. We introduce perturbations in both the metric tensor
\begin{equation}
\nu = \nu_0 + \delta \nu, \qquad \lambda = \lambda_0 + \delta \lambda,
\end{equation}
and the fluid variables
\begin{equation}
\rho = \rho_0 + \delta \rho, \qquad p_r = p_{r,0} + \delta p_r, \qquad p_t = p_{t,0} + \delta p_t,
\end{equation}
where the subscript $0$ labels the background solutions, and each perturbation $\delta A(t,r)$ is allowed to depend on both time and the radial coordinate. We then employ separation of variables, assuming a harmonic time dependence of the form $\delta A(t,r)=A(r) e^{i \omega t}$, where $\omega$ is the frequency to be determined. In the small-perturbation regime, the field equations and the conservation equations for the stress-energy tensor are linearized to first order, retaining only terms that are linear in the perturbations. Further details can be found, for example, in \cite{Dev:2003, Dev:2004, Arbanil:2016}.

\smallskip

The eigenfrequencies are obtained by solving the associated Sturm-Liouville boundary-value problem, subject to boundary conditions at the stellar centre, $r = 0$, and at the surface, $r = R$ \cite{Sagun:2020qvc}
\begin{align}
    \displaystyle \frac{\eta}{\xi} \biggm|_{r = 0} &= - 3 \Gamma (0) , \\
    \displaystyle \frac{\eta}{\xi} \biggm|_{r = R} &= \left( 1 - \frac{2 M}{R} \right)^{-1} \left( - \frac{M}{R} - \frac{\omega^2 R^3}{M} \right) - 4  .
\end{align}

Because $\Delta$ vanishes at both the centre and the stellar surface, these boundary conditions have the same form for isotropic and anisotropic configurations. For isotropic stars, they may be derived as follows \cite{chanmugam1}. At the centre, in the equation governing $\xi(r)$, the first term must remain finite in the limit $r \rightarrow 0$. This requirement gives $\eta(0)/\Gamma(0) + 3 \xi(0) =0$. At the surface, the pressure tends to zero as $r \rightarrow R$; hence, in the equation for $\eta$, the coefficient multiplying $(1/P(R))$ must vanish.

\medskip

\noindent It is useful to define the dimensionless quantity
\begin{equation}
s = \frac{\omega}{\omega_*}, \qquad \omega_* = \sqrt{\frac{M}{R^3}}.
\end{equation}
For fixed stellar mass and radius, integration of the perturbation equations from the centre to the surface yields eigenfunction values that depend only on the as-yet-undetermined frequency. The admissible values of $s$ are then found by solving the algebraic equation (50) graphically. The calculations were carried out with Wolfram Mathematica, using standard built-in routines such as NDSolve, Plot and FindRoot; no convergence difficulties were encountered.

\medskip

\noindent The oscillation frequencies are subsequently calculated as
\begin{equation}
\nu_n = \frac{\omega_n}{2\pi}=\frac{s_n}{2 \pi} \:\sqrt{\frac{M}{R^3}} = s_n \frac{\sqrt{M/R}}{2\pi} \: \frac{1}{R},
\end{equation}

\medskip

\noindent where $n=0,1,2,...$ denotes the number of nodes for a star with a specified mass and radius. The conversion from geometrical units to $kHz$ is performed using the conversion factors
\begin{equation}
1s = 3 \times 10^5~km, \qquad M_{\odot} \approx 1.5~km.
\end{equation}

\medskip

\noindent The so-called large frequency separation is a seismic quantity commonly employed in asteroseismology. It is defined as
\begin{equation}
    \Delta \nu_n = \nu_{n+1} - \nu_n, \; \; \; n=0,1,2,3, ...
\end{equation}

\medskip

\noindent Once the oscillation spectrum has been determined, this quantity can be evaluated directly. It represents the frequency difference between two successive modes and can be shown to depend on the stellar mass and radius \cite{tassoul, miglio, ilidio}.

A real frequency corresponds to purely oscillatory behaviour, with $\omega_n^2 > 0$. Conversely, if $\omega_n^2 < 0$, the frequency is imaginary and the associated perturbation grows exponentially with time. Negative values of $\omega_n^2$ therefore signal unstable radial oscillations. The condition $\omega_n^2=0$ consequently identifies the onset of radial instability, separating stable from unstable modes. It is well established that the fundamental frequency $\omega_0$ becomes imaginary when the central density exceeds the critical value at which the stellar mass, regarded as a function of $\rho_c$, reaches its maximum; see, for example, \cite{Horvat, Kokkotas:2000up, chanmugam2}.

\bigskip

\section{Results and discussion}

\hspace{0.5cm}

We have investigated the tidal Love numbers and deformabilities, together with the radial oscillation spectra, of relativistic stars composed of anisotropic quark matter. Our analysis considers three stellar masses: 0.7 $M_{\odot}$, 2.0 $M_{\odot}$, and 2.5 $M_{\odot}$. The corresponding radii are 9.04 $km$ (9.05 $km$), 12.13 $km$ (12.18 $km$), and 12.05 $km$ (12.40 $km$), respectively. For the radial oscillations, we calculate the 13 lowest modes, whose frequencies are reported in Tables \ref{tab:1} and \ref{tab:2}. For comparison, Table \ref{tab:3} presents the first two oscillation modes for the isotropic limit ($g = 0$), considering stellar masses of 0.7 $M_{\odot}$ and 2.0 $M_{\odot}$.


\begin{table}[ht!]
    \centering
    \begin{tabular}{|c|c|c|c|} \hline
       Mode order $n$ & 0.7 M$_{\odot}$ & 2.0 M$_{\odot}$ & 2.5 M$_{\odot}$  \\ \hline \hline
       0 & 7.77 & 2.84 & 0.29 \\ \hline
       1 & 16.89 & 7.93 & 5.12 \\ \hline
       2 & 25.69 & 12.41 & 8.33 \\ \hline
       3 & 34.42 & 16.78 & 11.40 \\ \hline
       4 & 43.12 & 21.11 & 14.41 \\ \hline
       5 & 51.81 & 25.42 & 17.40 \\ \hline
       6 & 60.48 & 29.72 & 20.37 \\ \hline
       7 & 69.16 & 34.01 & 23.33 \\ \hline
       8 & 77.83 & 38.30 & 26.29 \\ \hline
       9 & 86.49 & 42.58 & 29.25 \\ \hline
       10 & 95.16 & 46.86 & 32.20 \\ \hline
       11 & 103.82 & 51.14 & 35.15 \\ \hline
       12 & 112.49 & 55.42 & 38.09 \\ \hline
    \end{tabular}
    \caption{Case $g=0.8$: Frequencies (in kHz) of the radial oscillation for the 13 lowest modes considering three different stellar masses: 0.7 $M_{\odot}$, 2.0 $M_{\odot}$, and 2.5 $M_{\odot}$.}
    \label{tab:1}
\end{table}


\begin{table}[ht!]
    \centering
    \begin{tabular}{|c|c|c|c|} \hline
       Mode order $n$ & 0.7 M$_{\odot}$ & 2.0 M$_{\odot}$ & 2.5 M$_{\odot}$  \\ \hline \hline
       0 & 7.78 & 2.90 & 1.21 \\ \hline
       1 & 16.89 & 7.99 & 5.57 \\ \hline
       2 & 25.69 & 12.48 & 8.94 \\ \hline
       3 & 34.42 & 16.87 & 12.18 \\ \hline
       4 & 43.12 & 21.22 & 15.38 \\ \hline
       5 & 51.81 & 25.55 & 18.55 \\ \hline
       6 & 60.49 & 29.87 & 21.70 \\ \hline
       7 & 69.16 & 34.18 & 24.85 \\ \hline
       8 & 77.83 & 38.49 & 27.00 \\ \hline
       9 & 86.50 & 42.79 & 31.14 \\ \hline
       10 & 95.16 & 47.10 & 34.28 \\ \hline
       11 & 103.83 & 51.39 & 37.41 \\ \hline
       12 & 112.49 & 55.69 & 40.55 \\ \hline
    \end{tabular}
    \caption{Case $g=1.0$: Frequencies (in kHz) of the radial oscillation for the 13 lowest modes considering three different stellar masses: 0.7 $M_{\odot}$, 2.0 $M_{\odot}$, and 2.5 $M_{\odot}$.}
    \label{tab:2}
\end{table}


\begin{table}[ht!]
    \centering
    \begin{tabular}{|c|c|c|} \hline
       Mode order $n$ & 0.7 M$_{\odot}$ & 2.0 M$_{\odot}$ \\ \hline \hline
       0 & 7.75 & 2.45 \\ \hline
       1 & 16.87 & 7.56 \\ \hline
    \end{tabular}
    \caption{Case $g=0.0$: Frequencies (in kHz) of the radial oscillation for the fundamental and first excited modes, considering the stellar masses 0.7 $M_{\odot}$ and 2.0 $M_{\odot}$.}
    \label{tab:3}
\end{table}

\begin{figure}[h]
    \centering
    \includegraphics[scale=1]{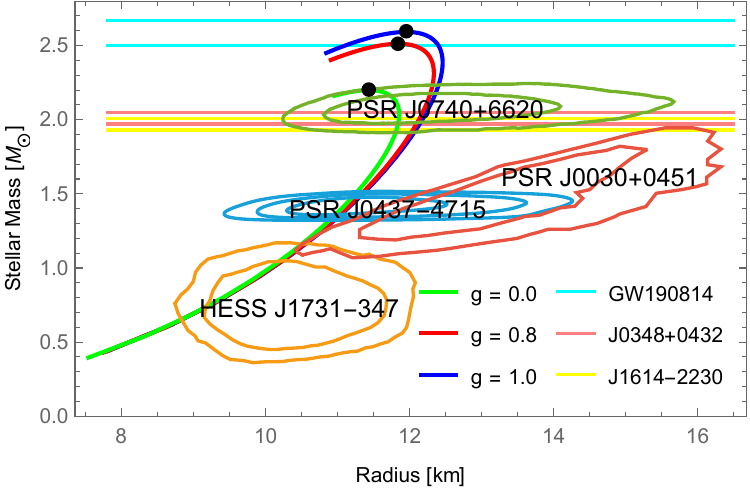} 
    \caption{
    Mass-to-radius profile with non-vanishing anisotropy (in green for $g = 0.0$, in red for $g = 0.8$ and blue for $g = 1.0$) of the CFL model we have chosen from \cite{VasquezFlores:2017uor} (CFL10). The black dots represent the maximum mass value for each profile: $2.59 \: M_{\odot}$ for $g = 1.0$, $2.51 \: M_{\odot}$ for $g = 0.8$, and $2.20 \: M_{\odot}$ for $g = 0.0$. We have added the contours and mass ranges of some possible strange star candidates. These include the contours of the supernova remnant HESS J1731-347 (orange), and the pulsars PSR J0437-4715 (light blue), PSR J0030+0451 (red), and PSR J0740+6620 (green). The profile passes through these contours and also through the mass ranges of the pulsars PSR J0348+0432 (pink) and PSR J1614-2230 (yellow), and of the higher-mass gravitational-wave signal {\bf GW}190814 (cyan). The part of the profile after the highest stellar mass corresponds to unstable configurations.
    }
    \label{fig:profile}
\end{figure}

Figure \ref{fig:profile} displays the mass-radius relation, together with several current astronomical constraints. The resulting mass-radius sequence for CFL quark matter with a positive anisotropy parameter is compatible with a range of observational data, including the massive pulsars PSR J0740+6620, PSR J0348+0432 and PSR J1614-2230, as well as the gravitational-wave event {\bf GW}190814. These comparisons indicate that the model considered here is viable, although it is certainly not unique. The principal CFL EoS parameters, such as the superconducting gap and the bag constant, are subject to substantial theoretical uncertainties; consequently, other choices of these parameters may also be consistent with the observational constraints. The segment of the M-R curve beyond the maximum-mass configuration represents unstable stellar solutions, as discussed below in connection with the Harrison-Zel’dovich stability criterion.

\smallskip

Figure \ref{fig:lambda} presents the dimensionless tidal deformability, $\Lambda$, as a function of the stellar mass. The blue vertical segment marks the observationally allowed interval for the deformability of a canonical star of mass 1.4 $M_{\odot}$, namely $\Lambda_{1.4} \: = \: 190^{+390}_{-120}$ \cite{LIGOScientific:2018cki}. 

We also examined additional CFL models, considering both isotropic and anisotropic configurations; however, none of these alternatives satisfied all the astrophysical constraints adopted in this study. The nature of the secondary component in the event {\bf GW}190814 remains uncertain. Thus, requiring the model to support a star with a mass of approximately $2.5~M_{\odot}$ is conditional upon interpreting this object as a quark star. Although a sufficiently stiff EoS can support such a high mass even in the absence of anisotropy, it leads to an excessively large dimensionless tidal deformability. This provides a separate motivation for considering the anisotropic model adopted here. For the CFL10 model studied in this work, a small value of $g$ is insufficient to support a $2.5~M_{\odot}$ configuration, thereby motivating our choice of the anisotropy-strength parameter. We therefore compare the cases $g=0.8, 1.0$. While their numerical predictions exhibit small differences, the main conclusions remain the same.

\smallskip

The comparison of our predictions with the tidal-deformability constraint inferred from {\bf GW}170817 should nevertheless be regarded with caution. The often-quoted bound on $\Lambda_{1.4}$ relies on particular assumptions about the binary components and their shared EoS; its derivation may also employ EoS-insensitive relations calibrated mainly for standard neutron-star configurations. The objects investigated here are anisotropic, self-bound quark stars, whose tidal response need not coincide with that of isotropic neutron stars. More specifically, anisotropy affects both the equilibrium structure and the perturbation equations, whereas the non-zero surface energy density calls for suitable matching conditions for the tidal perturbations. Recent works have examined the validity of universal relations for anisotropic quark stars (see, for example, \cite{Pretel:2024pem}), but their applicability still depends on the stellar model and on the adopted prescription for anisotropy. Accordingly, the comparison with {\bf GW}170817 should be viewed as an indicative test of the observational viability of our configurations, rather than as a definitive constraint. A more conclusive assessment would require a dedicated Bayesian analysis based on gravitational-wave templates that consistently account for the anisotropic tidal response of the binary components.

\smallskip

\begin{figure}[b]
\centering
\includegraphics[scale=1.15]{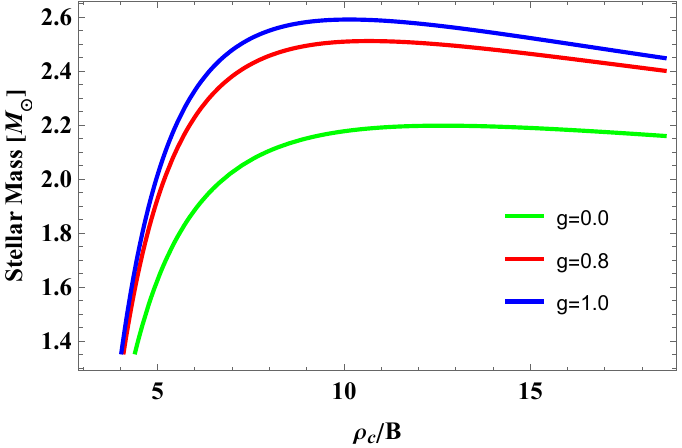}
\caption{
Stellar mass (in solar masses) versus dimensionless central energy density (in units of the bag constant) for the two anisotropic models discussed in this work. The red curve corresponds to $g=0.8$, while the blue curve to $g=1.0$. According to the Harrison-Zel'dovich criterion, only the increasing part of the curves implies stability. For the sake of comparison, the case $g=0.0$ is included as well.
}
\label{fig:0}
\end{figure}

Figure \ref{fig:0} shows how the stellar mass varies with the central energy density. According to the Harrison-Zel’dovich criterion \cite{Harrison,Zeldovich},
\begin{equation}
\frac{dM}{d \rho_c} > 0, \; \; \; \rightarrow  \; \; \; \textrm{stability},
\end{equation}
\begin{equation}  
\frac{dM}{d \rho_c} < 0, \; \; \; \rightarrow  \; \; \; \textrm{instability},
\end{equation}
stable configurations are therefore confined to the increasing branches of the curves. The maximum mass identified in this analysis coincides with the upper mass endpoint of the M-R sequences and with the critical mass at which radial stability is lost, namely the point where $\nu_0^2=0$ \cite{Horvat, Kokkotas:2000up, chanmugam2}.

\smallskip


\smallskip

\begin{figure}[t]
    \centering
    \includegraphics[scale=1.05]{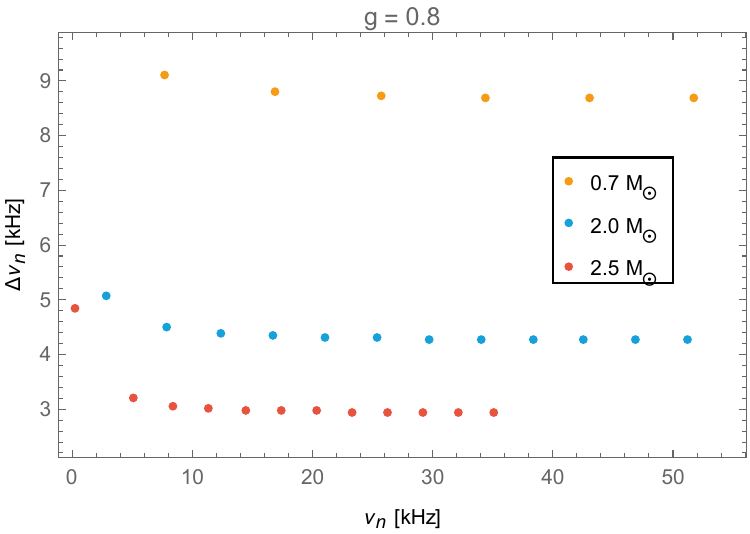} \
    \caption{
    Case $g=0.8$: Large frequency separations $\Delta \nu_{n}$ in kHz, for 0.7 $M_{\odot}$ in yellow, 2.0 $M_{\odot}$ in blue, and 2.5 $M_{\odot}$ in red. We consider the first 13 oscillation modes. Due to the high values of the frequencies for 0.7 $M_{\odot}$, only the first 6 frequencies are displayed, instead of 13. For all values, see Table \ref{tab:1}.}
    \label{fig:frecuencies08}
    \end{figure}

\begin{figure}[b]
    \centering
    \includegraphics[scale=1.05]{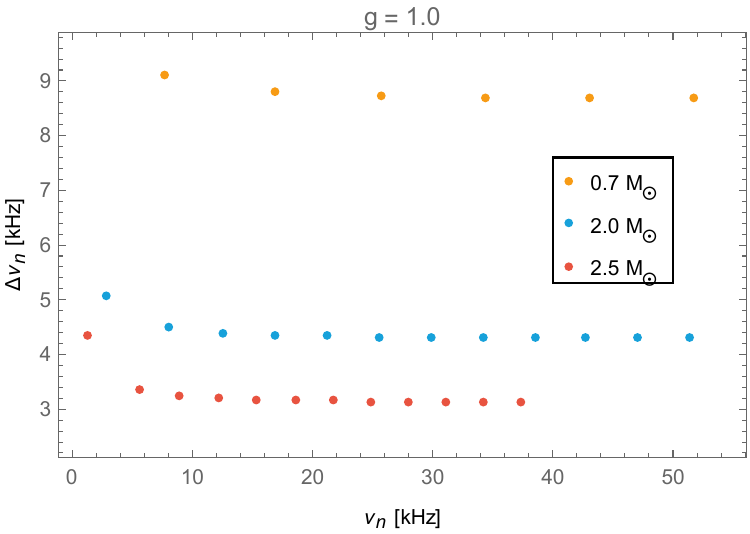} \
    \caption{
    Case $g=1.0$: Large frequency separations $\Delta \nu_{n}$ in kHz, for 0.7 $M_{\odot}$ in yellow, 2.0 $M_{\odot}$ in blue, and 2.5 $M_{\odot}$ in red. We consider the first 13 oscillation modes. Due to the high values of the frequencies for 0.7 $M_{\odot}$, only the first 6 frequencies are displayed, instead of 13. For all values, see Table \ref{tab:2}.}
    \label{fig:frecuencies1}
    \end{figure}

The large frequency separations, $\Delta \nu_{n}$ (kHz), are displayed in Figures \ref{fig:frecuencies08} and \ref{fig:frecuencies1}. Our results indicate that the lowest-mass star has the highest sound speed and the largest frequency separation. Conversely, the most massive star exhibits the lowest sound speed and the smallest large frequency separation. The oscillation frequencies for a star of mass 0.7 $M_{\odot}$ are also higher than those obtained for the 2.0 $M_{\odot}$ and 2.5 $M_{\odot}$ configurations. Indeed, only approximately half of the 13 modes calculated in this work can be displayed in Figures \ref{fig:frecuencies08} and \ref{fig:frecuencies1}. 

\smallskip

\begin{figure}[p]
    \centering
    \includegraphics[scale=1.05]{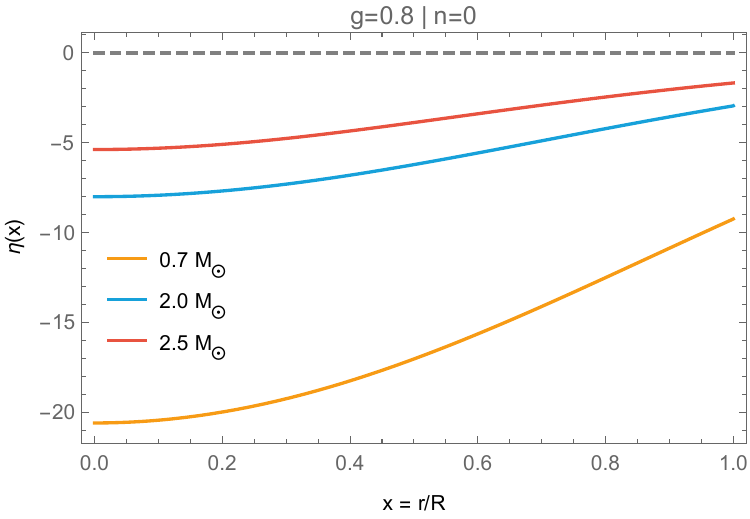} \\
    \vspace{0.5cm}
    \includegraphics[scale=1.05]{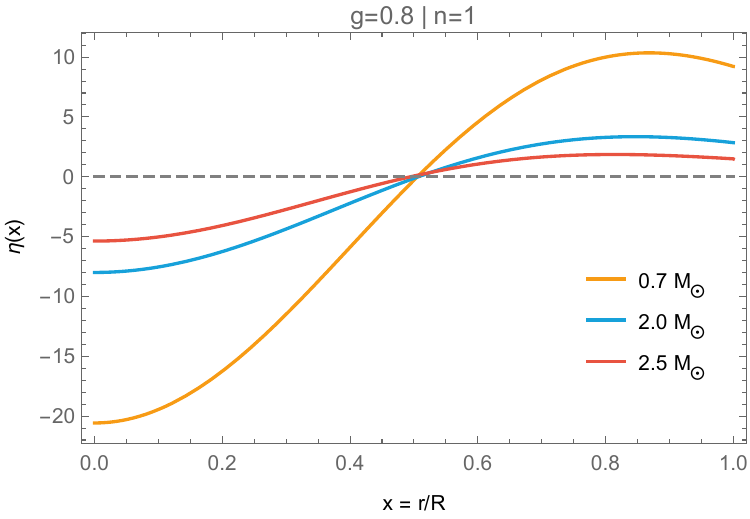} \\
    \vspace{0.5cm}
    \caption{
    Case $g=0.8$: Radial profiles of the perturbation $\eta(x)$ versus dimensionless radial coordinate, $r/R$, of the fundamental mode $n = 0$ (up) and first excited state $n = 1$ (down), for 0.7 $M_{\odot}$ in yellow, 2.0 $M_{\odot}$ in blue, and 2.5 $M_{\odot}$ in red.}
    \label{fig:eta08}
\end{figure}

\begin{figure}[p]
    \centering
    \includegraphics[scale=1.05]{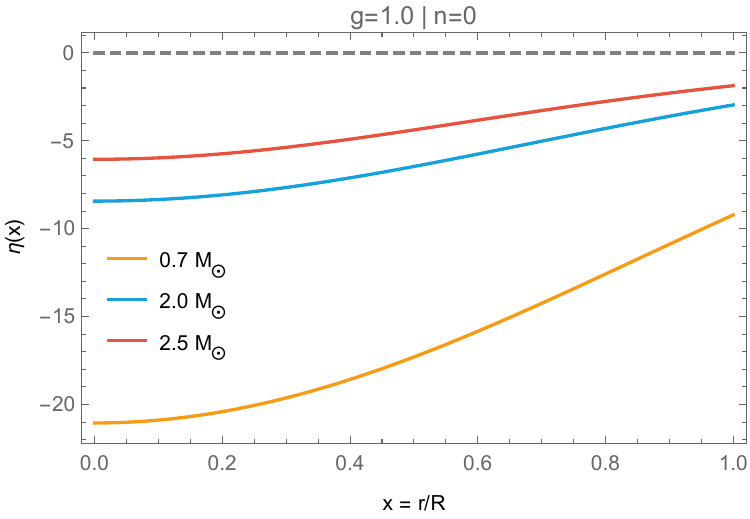} \\
    \vspace{0.5cm}
    \includegraphics[scale=1.05]{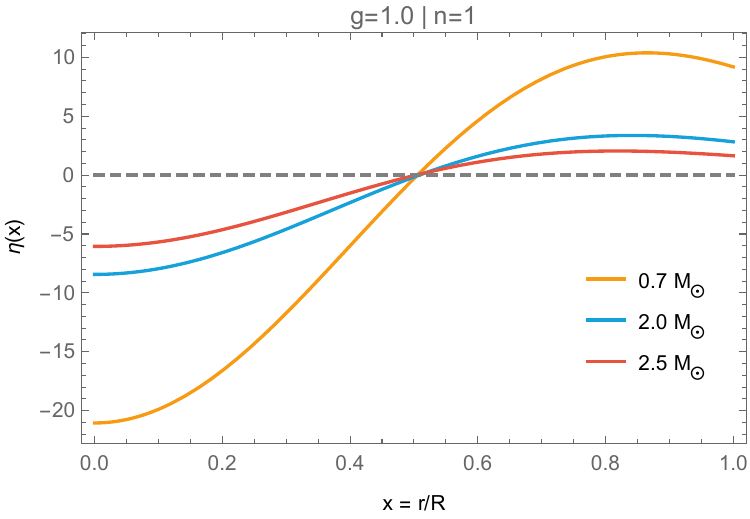} \\
    \vspace{0.5cm}
    \caption{
    Case $g=1.0$: Radial profiles of the perturbation $\eta(x)$ versus dimensionless radial coordinate, $r/R$, of the fundamental mode $n = 0$ (up) and first excited state $n = 1$ (down), for 0.7 $M_{\odot}$ in yellow, 2.0 $M_{\odot}$ in blue, and 2.5 $M_{\odot}$ in red.}
    \label{fig:eta1}
\end{figure}

\begin{figure}[p]
    \centering
    \includegraphics[scale=1.05]{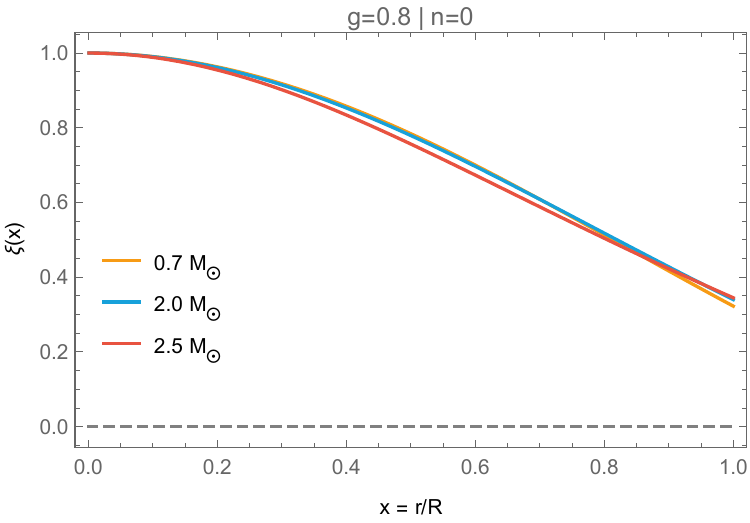} \\
    \vspace{0.5cm}
    \includegraphics[scale=1.05]{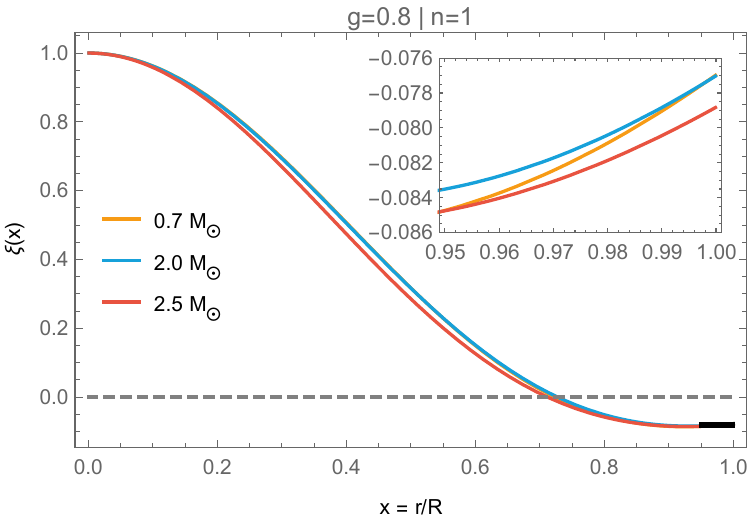} \\
    \vspace{0.5cm}
    \caption{
     Case $g=0.8$: Perturbation $\xi (x)$ versus dimensionless radial coordinate $r/R$, of the fundamental mode $n = 0$ (up) and first excited state $n = 1$ (down), for 0.7 $M_{\odot}$ in yellow, 2.0 $M_{\odot}$ in blue, and 2.5 $M_{\odot}$ in red.}
    \label{fig:xi08}
\end{figure}

\begin{figure}[p]
    \centering
    \includegraphics[scale=1.05]{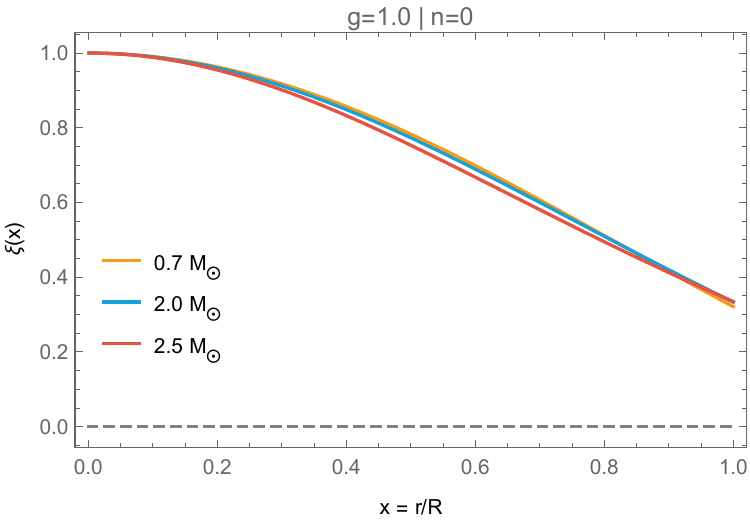} \\
    \vspace{0.5cm}
    \includegraphics[scale=1.05]{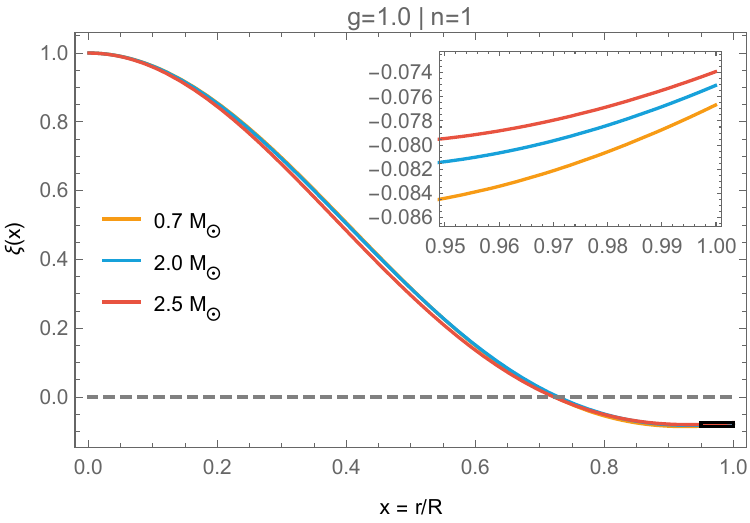} \\
    \vspace{0.5cm}
    \caption{
     Case $g=1.0$: Perturbation $\xi (x)$ versus dimensionless radial coordinate $r/R$, of the fundamental mode $n = 0$ (up) and first excited state $n = 1$ (down), for 0.7 $M_{\odot}$ in yellow, 2.0 $M_{\odot}$ in blue, and 2.5 $M_{\odot}$ in red.}
    \label{fig:xi1}
\end{figure}

Figures \ref{fig:eta08}, \ref{fig:eta1}, \ref{fig:xi08} and \ref{fig:xi1} show the fundamental ($n=0$) and first excited ($n=1$) modes of the two perturbation functions, $\eta (x)$ and $\xi (x)$, for each of the three stellar masses. The $\xi$ profiles exhibit only small differences among the three masses for both modes. For the fundamental mode, all these curves begin at $1$ and decrease monotonically. By contrast, the dependence of the $\eta$ profiles on stellar mass is much more pronounced. In the $n=0$ case, all three curves increase monotonically, with the profile for the most massive star lying above the other two.

\smallskip

\begin{figure}[p]
    \centering
    \textcolor{gray}{\Large \hspace{0.8cm} $g = 0.8$} \par \medskip
    \includegraphics[scale=0.5]{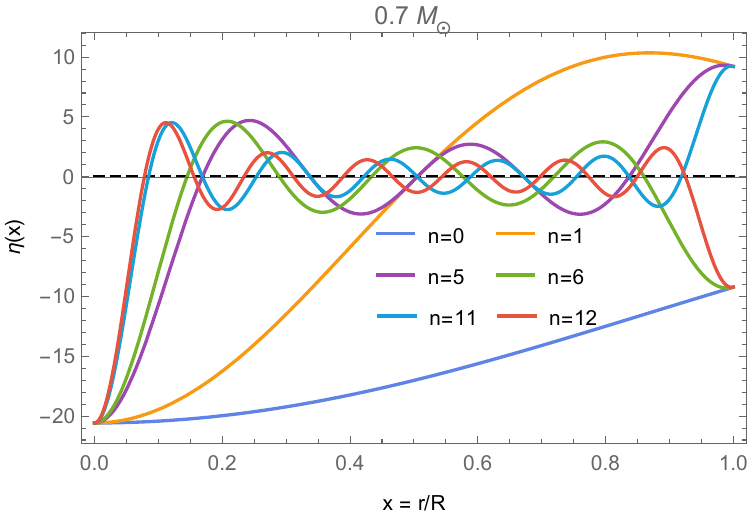} \
    \includegraphics[scale=0.5]{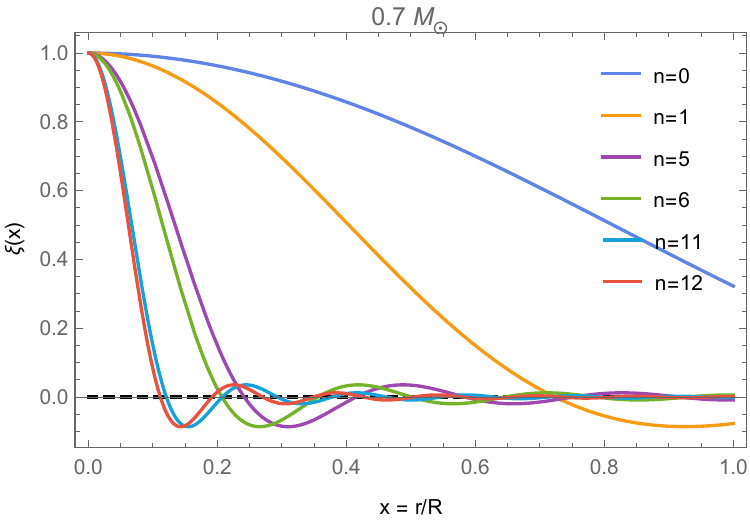} \\
    \vspace{0.3cm}
    \includegraphics[scale=0.5]{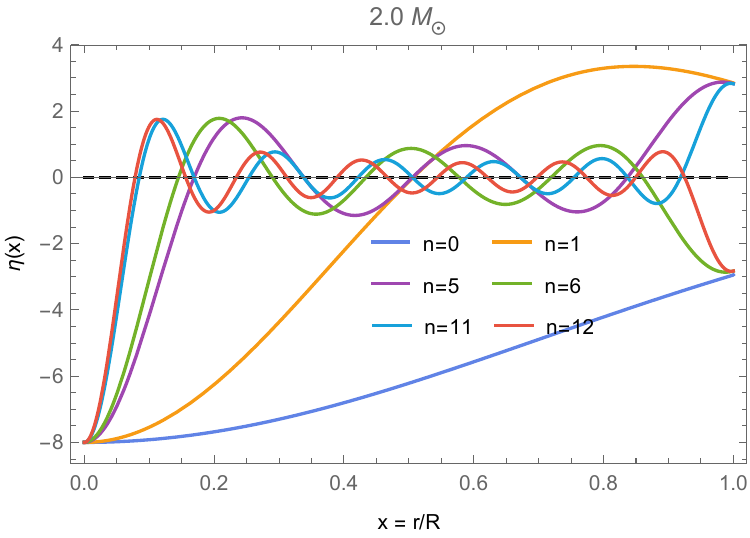} \
    \includegraphics[scale=0.5]{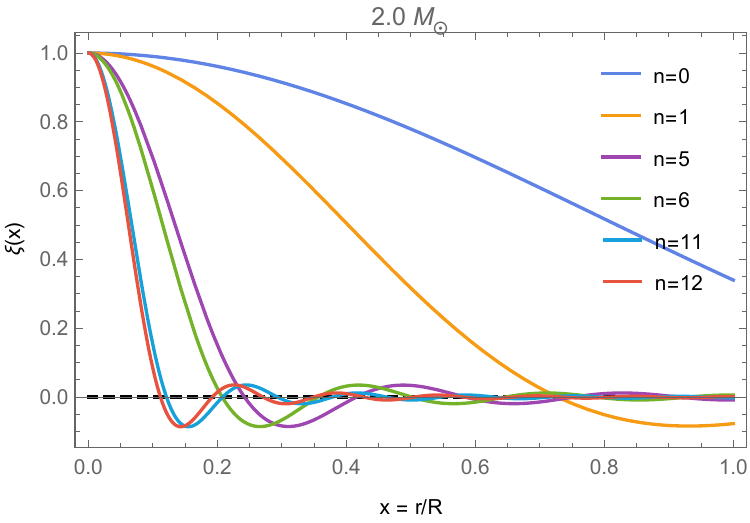} \\
    \vspace{0.3cm}
    \includegraphics[scale=0.5]{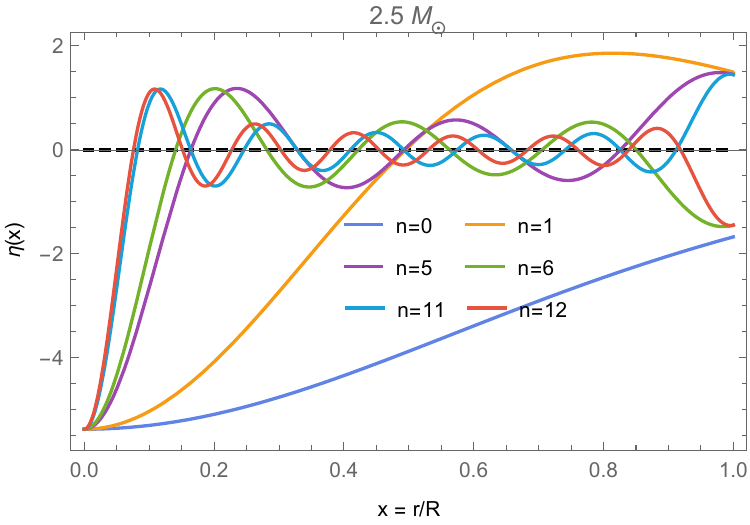} \
    \includegraphics[scale=0.5]{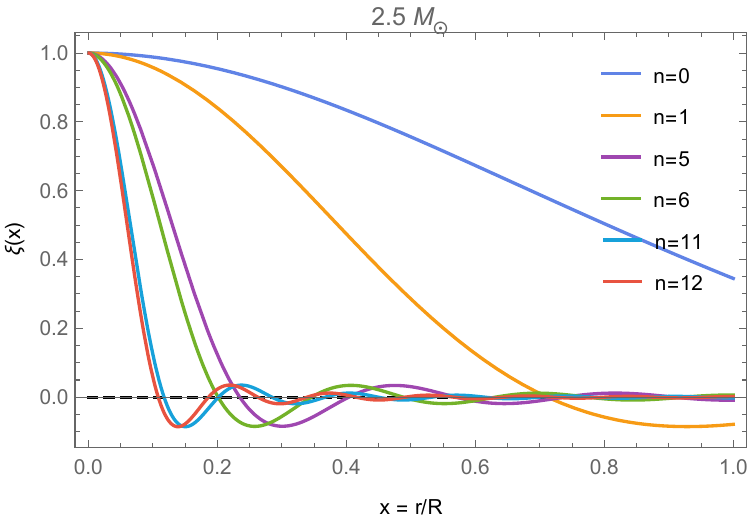} \\
    \vspace{0.3cm}
    \caption{
    Case $g=0.8$: Radial oscillations of the perturbations $\eta (x)$ and $\xi (x)$ of the three stars, as a function of the normalized radial coordinate $r/R$, for the modes $n$ = 0 (blue), 1 (orange), 5 (purple), 6 (green), 11 (light blue) and 12 (red).}
    \label{fig:eta&xi08}
\end{figure}

\begin{figure}[p]
    \centering
    \textcolor{gray}{\Large \hspace{0.8cm} $g = 1.0$} \par \medskip
    \includegraphics[scale=0.5]{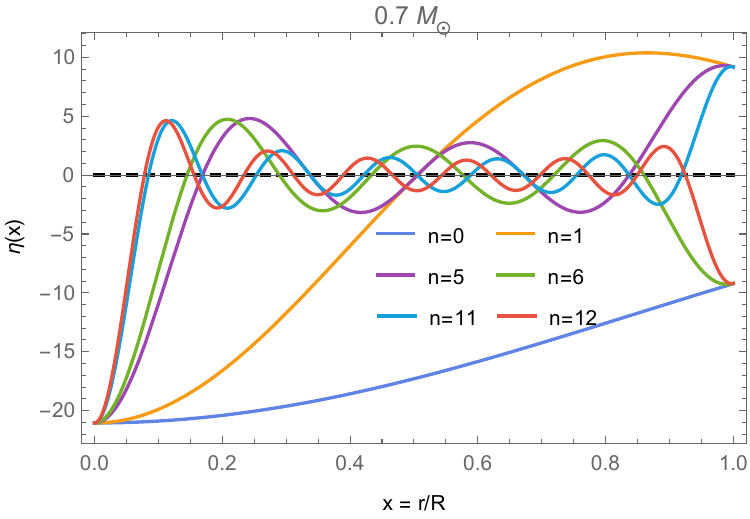} \
    \includegraphics[scale=0.5]{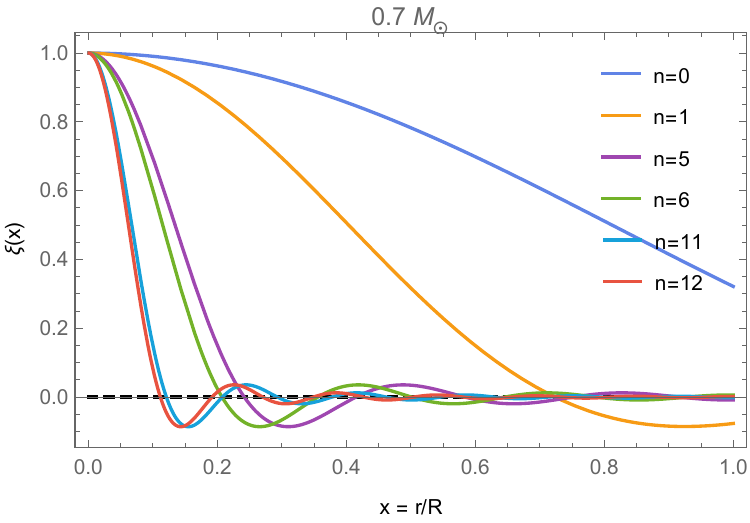} \\
    \vspace{0.3cm}
    \includegraphics[scale=0.5]{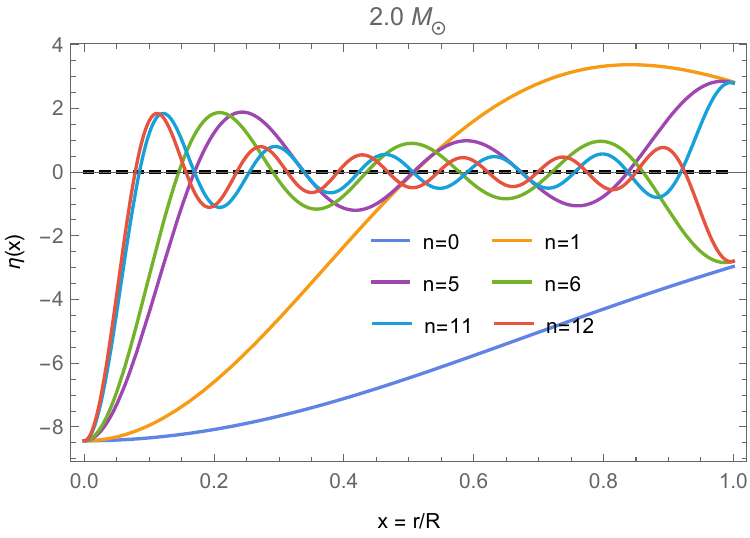} \
    \includegraphics[scale=0.5]{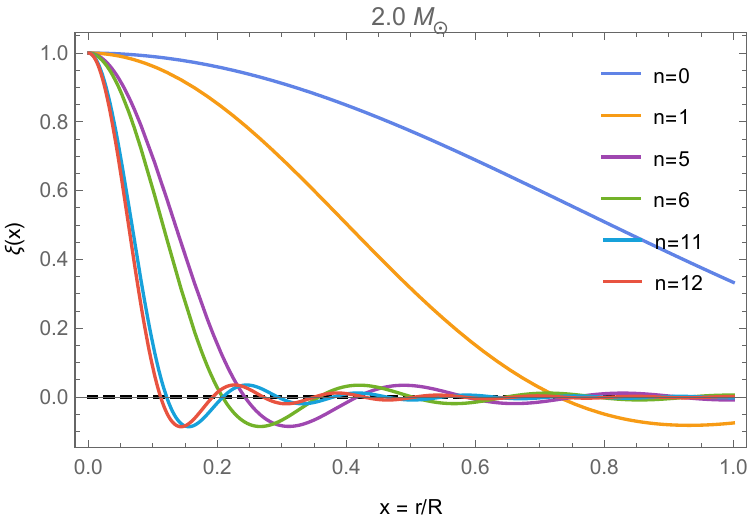} \\
    \vspace{0.3cm}
    \includegraphics[scale=0.5]{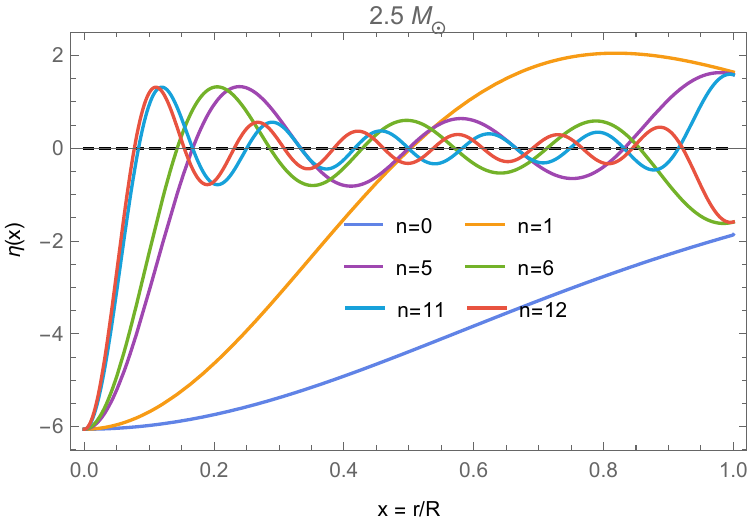} \
    \includegraphics[scale=0.5]{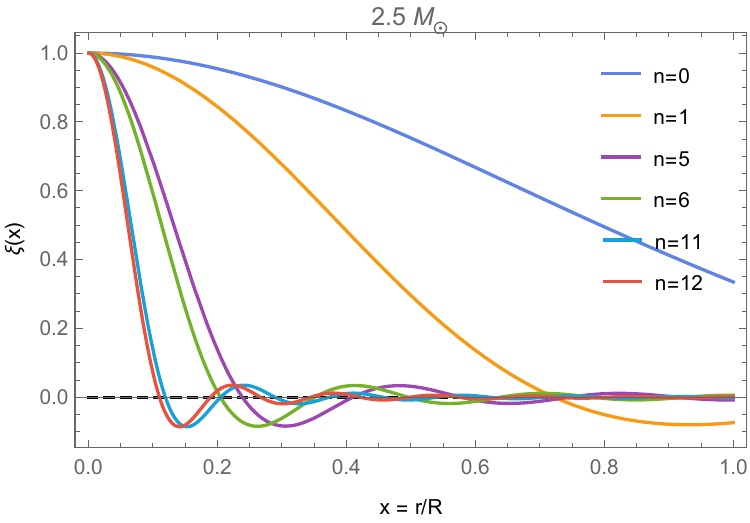} \\
    \vspace{0.3cm}
    \caption{
    Case $g=1.0$: Radial oscillations of the perturbations $\eta (x)$ and $\xi (x)$ of the three stars, as a function of the normalized radial coordinate $r/R$, for the modes $n$ = 0 (blue), 1 (orange), 5 (purple), 6 (green), 11 (light blue) and 12 (red).}
    \label{fig:eta&xi1}
\end{figure}

The radial behaviour of the perturbations $\eta (x)$ and $\xi (x)$ for the three stellar configurations is presented in Figures \ref{fig:eta&xi08} and \ref{fig:eta&xi1}, plotted against the dimensionless radial coordinate $r/R$. The displayed modes are $n =$ 0, 1, 5, 6, 11 and 12. As expected, the number of nodes of each eigenfunction agrees with the corresponding mode number.

\smallskip

\begin{figure}[t]
    \centering 
    \includegraphics[scale=1]{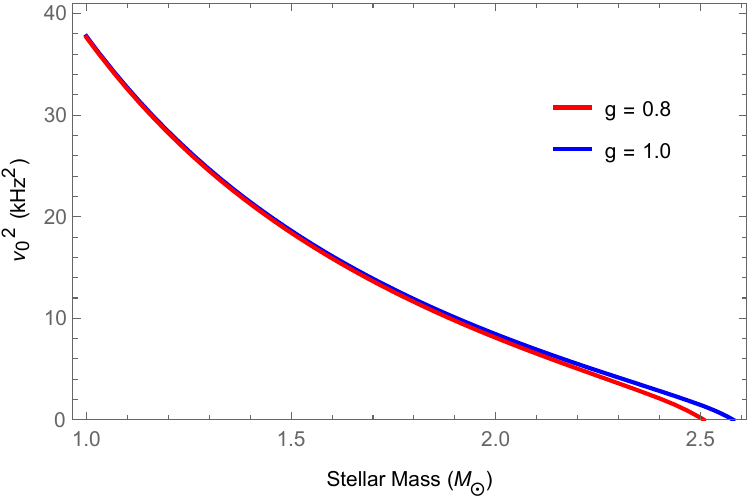}
    \caption{Comparison between $g=0.8$ and $g=1$ for stellar mass (M$_{\odot}$) vs he squared frequency of the fundamental mode (kHz$^2$).}
    \label{fig:massfrequency}
\end{figure}

Finally, Figure~\ref{fig:massfrequency} illustrates that the vanishing of the squared fundamental-mode frequency, $\nu_0^2 = 0$, occurs precisely at the maximum mass of each model: $2.51\,M_{\odot}$ for $g = 0.8$ and $2.59\,M_{\odot}$ for $g = 1.0$. These values agree with the maximum masses obtained from the mass-radius sequences in Figure~\ref{fig:profile}.

\smallskip


\smallskip

The subject of this study is related to that of several earlier works cited in the bibliography, all of which examined relativistic stellar configurations in GR without a cosmological constant. In addition, the stars considered in those studies were electrically neutral and non-rotating. Nevertheless, there are important distinctions between those analyses and the present one. The study in \cite{Arbanil}, for example, considered polytropic EoSs and adopted negative anisotropy. In \cite{Sepulveda:2024hdc}, the stellar matter consisted of two dark-sector fluid components, but anisotropy was not included. Finally, \cite{Panotopoulos:2025ixx} constructed mass-radius sequences for quark stars admixed with dark matter within the two-fluid formalism; however, neither radial oscillations nor tidal Love numbers were investigated.

Before closing, we would like to make one further remark. Distinguishing quark stars from conventional neutron stars is an interesting and important problem. In this respect, \cite{Fukushima:2025dna} demonstrated that magnetic-type tidal Love numbers provide a robust means of discriminating between quark stars and neutron stars, whereas electric-type Love numbers alone are not sufficient when the two classes of objects have comparable masses and radii. We leave this question for future work and hope to investigate it in a forthcoming continuation of the present study.


\section{Conclusions}

\hspace{0.5cm}

In this work, we have investigated the gravito-electric tidal Love numbers and radial oscillation modes of fluid spheres composed of anisotropic quark matter. We adopted the analytic CFL equation of state together with a prescribed ansatz for the anisotropy factor. We began by constructing interior configurations in hydrostatic equilibrium: the TOV equations were integrated numerically, allowing us to obtain the mass-radius relation. We then integrated the Riccati equation governing the even-parity metric perturbations and thereby determined the gravito-electric tidal Love numbers, together with the associated dimensionless tidal deformability as a function of stellar mass. As illustrated by our results, the model considered here is compatible with several current astronomical constraints, including those associated with the most massive pulsars, the HESS compact object, NICER observations, and the {\bf GW}170817 and {\bf GW}190814 events.

\smallskip

We subsequently examined the radial oscillations of pulsating stars. For three different stellar masses, we solved the perturbation equations numerically, imposing the appropriate boundary conditions at the stellar centre and surface, in order to determine the permitted frequencies and their corresponding eigenfunctions. We also calculated the large frequency separations. Our results indicate that these separations approach a constant value for higher excited modes, as expected from the asymptotic theory. The frequency values are reported in Table \ref{tab:1} and Table \ref{tab:2}, while the behaviour of the relevant quantities is illustrated in several figures.

\smallskip

The radial profiles of the perturbations display the characteristic behaviour expected for Sturm-Liouville problems: the fundamental modes have no nodes, the first excited modes have one node, and so forth. Our results further indicate that the spectrum of the least massive star is associated with higher frequencies, whereas the most massive configuration considered here exhibits the lowest frequencies. This trend is consistent with the sound speed being higher throughout the less massive star and lower throughout the most massive one.

\smallskip

Finally, we compared the radial profiles of the perturbations for the fundamental and first excited modes across all three models. For the $\xi$ eigenfunctions, the differences between the models are more pronounced in the $n=0$ case: the eigenfunction associated with the $0.7~M_{\odot}$ star decreases more rapidly, while that corresponding to the $2.5~M_{\odot}$ star decreases more gradually as the radial coordinate increases across the stellar interior. For the $\eta$ eigenfunctions, by contrast, appreciable differences between the models are present in both the $n=0$ and $n=1$ cases. In particular, the functions associated with the lowest-mass configuration take a more negative value at the origin, owing to its higher relativistic adiabatic index.

\section*{Funding:}
\noindent
This research received no external funding.

\section*{Data Availability Statement:}
\noindent
The raw data supporting the conclusions of this article will be made available by the authors on request.

\section*{Conflicts of Interest:}
\noindent
The authors declare no conflicts of interest.

\section*{Acknowledgements:}
\noindent
The authors thank the anonymous reviewers for useful comments and suggestions.

\newpage



\end{document}